\documentclass[11pt]{article}

\usepackage[top=1in, bottom=1in, left=1in, right=1in]{geometry}

\usepackage{amsmath}
\usepackage{amsfonts}
\usepackage{wasysym}
\usepackage{graphicx}
\usepackage{comment}

\usepackage[compress,square,numbers]{natbib}

\usepackage{hyperref}
\numberwithin{equation}{section}

\def\tn{\textnormal}

\begin{document}

%---TITLE PAGE------------------------------------------------------------------
%\begin{comment}
\thispagestyle{empty}

\hfill March 29, 2010

\addvspace{45pt}

\begin{center}

\Large{\textbf{The Effective K\"ahler Potential, Metastable Vacua and R-Symmetry Breaking in O'Raifeartaigh Models}}
\\[35pt]
\large{Shermane Benjamin, Christopher Freund and Ben Kain}
\\[10pt]
\textit{Department of Physics and Astronomy}
\\ \textit{Rowan University}
\\ \textit{201 Mullica Hill Road, Glassboro, NJ 08028, USA}
\end{center}

\setcounter{footnote}{0}

\addvspace{35pt}

\begin{abstract}
\noindent Much has been learned about metastable vacua and R-symmetry breaking in O'Raifeartaigh models.  Such work has largely been done from the perspective of the superpotential and by including Coleman-Weinberg corrections to the scalar potential.  Instead, we consider these ideas from the perspective of the one loop effective K\"ahler potential.  We translate known ideas to this framework and construct convenient formulas for computing individual terms in the expanded effective K\"ahler potential.  We do so for arbitrary R-charge assignments and allow for small R-symmetry violating terms so that both spontaneous and explicit R-symmetry breaking is allowed in our analysis.
\end{abstract} 

\newpage
%\end{comment}

%\tableofcontents
%\newpage
%===================================================================================================

\section{Introduction}
\label{intro}

Supersymmetry is a strong candidate for physics beyond the standard model to be seen at the LHC.  Inarguably, it is the most studied candidate and much has been said in the way of how it can be broken.  Such work is largely devoted to building models with nonzero vacuum energy in a global minimum of the potential.  Intriligator, Seiberg and Shih revitalized the idea that our universe could instead be in a long lived, local minimum of the potential, i.e. a metastable vacuum \cite{esr, ISS1, ISrev}.  They gave simple arguments, based on R-symmetry, that we are in fact led to this possibility \cite{ISS2, ISrev}.

Metastable vacua and R-symmetry can be studied in renormalizable, perturbative superpotentials of the O'Raifeartaigh type \cite{ISS1, ISrev, ISS2, ray, Shih, ferretti, ako, aldmar, dienes, sun, ms, sun2, kitano0, amariti, KomShih, Shih2}.  We will focus on O'Raifeartaigh superpotentials of the form \cite{Shih}
\begin{equation} \label{orsp1}
	W=fX + \frac{1}{2}\left(M_{ij}+XN_{ij}\right)\phi_i\phi_j + \cdots,
\end{equation}
where the dots denote possible cubic interactions of the $\phi_i$ fields.  At tree level, supersymmetry breaking extrema exist at $\phi_i=0$ and arbitrary $X$.  We will assume these are minima (which occurs when the mass matrix for the $\phi_i$ fields has no negative eigenvalues), so that we have a degenerate pseudomoduli space of supersymmetry breaking vacua, parametrized by $X$, the pseudomodulus \cite{ray,sun}.  There may or may not exist supersymmetric vacua elsewhere \cite{ferretti}.

Nelson and Seiberg \cite{ns} have shown that in a generic model of $F$-term supersymmetry breaking R-symmetry is a necessary condition.  On the other hand, Majorana gaugino masses, necessary in realistic models, require broken R-symmetry.  R-symmetry may be broken spontaneously or explicitly.  Spontaneous R-symmetry breaking in the superpotential (\ref{orsp1}) was analyzed by Shih \cite{Shih}.  He took (\ref{orsp1}) to be R-symmetric, satisfying selection rules
\begin{equation} \label{srules}
	M_{ij}\neq0 \Rightarrow R(\phi_i)+R(\phi_j)=2, \qquad N_{ij}\neq 0 \Rightarrow R(\phi_i)+R(\phi_j)=0,
\end{equation}
and showed that spontaneous R-symmetry breaking can occur if there exist fields with R-charge assignments different from zero and two.  Intriligator, Seiberg and Shih considered the possibility of explicit R-symmetry breaking through the introduction of small R-symmetry violating terms \cite{ISS2} and Marqu\'es and Schaposnik \cite{ms} analyzed their inclusion in the superpotential (\ref{orsp1}) (see also \cite{ako}).  From Nelson and Seiberg's argument, the introduction of R-symmetry violating terms establishes a supersymmetric vacuum.  The supersymmetric vacuum must vanish in the limit that the R-symmetry violating terms vanish, and therefore, for very small R-symmetry violating terms, is far from the origin.  Since R-symmetry is broken, but only slightly, the supersymmetry breaking vacuum should persist and we can expect it to be close to, but not at, the origin.  The supersymmetry breaking vacuum is then a long lived, metastable vacuum with broken R-symmetry \cite{ISS2}.

Including small, R-symmetry violating terms generalizes the O'Raifeartaigh superpotential (\ref{orsp1}) to
\begin{equation} \label{orsp2}
	W=fX + \frac{1}{2}\epsilon_X X^2 + \frac{1}{2}\left(M_{ij}+XN_{ij}\right)\phi_i\phi_j+\cdots.
\end{equation}
Since the explicit R-symmetry breaking is small, we require $\epsilon_X \ll 1$.  Including R-symmetry violating terms means the superpotential (\ref{orsp2}) no longer satisfies selection rules (\ref{srules}), because now $M_{ij}$ and $N_{ij}$ may contain small, R-symmetry violating terms.

Loop corrections play an important role in analyzing the vacuum structure of the theory.  As previously mentioned, there exists a degenerate pseudomoduli space of supersymmetry breaking vacua, parametrized by $X$, at $\phi_i=0$.  If loop corrections are included this degeneracy is lifted.  The full one loop, Coleman-Weinberg correction to the potential is given by \cite{colwei}
\begin{equation} \label{cw}
	V_{\tn{1-loop}} = \frac{1}{64\pi^2}\tn{Str}\left[ \mathcal{M}^4 \ln \left(\frac{\mathcal{M}^2}{\Lambda^2}\right)\right],
\end{equation}
where
\begin{equation} \label{massmatrix}
	\mathcal{M}_{ij} = M_{ij}+XN_{ij}
\end{equation}
is the mass matrix for the $\phi_i$ fields in (\ref{orsp1}) and (\ref{orsp2}) and $\Lambda$ is a momentum cutoff.  Loop corrections may instead be included through the effective K\"ahler potential \cite{effK, ISS1},
\begin{equation} \label{effk}
	K_{\tn{eff}} = K_{\tn{tree}} + K_{\tn{1-loop}},
\end{equation}
where we take $K_{\tn{tree}}=|X|^2$ to have minimal form and
\begin{equation} \label{kp1l}
	K_{\tn{1-loop}} = -\frac{1}{32\pi^2}\tn{Tr}\left[\mathcal{M}^\dagger \mathcal{M}\ln\left(\frac{\mathcal{M}^\dagger \mathcal{M}}{\Lambda^2}\right)\right].
\end{equation}
If the supersymmetry breaking is small the effective K\"ahler potential can be used to find the one loop corrected scalar potential \cite{ISS1}, as an alternative to the Coleman-Weinberg formula (\ref{cw}), through
\begin{equation}\label{spform}
	V = V_{\tn{tree}} + V_{\tn{1-loop}} = K_{\tn{eff}}^{X\overline{X}} |W_X|^2,
\end{equation}
where $K^{X\overline X}_{\tn{eff}}$ is the inverse effective K\"ahler metric and, since our interest is in $\phi_i=0$ vacua,
\begin{equation} \label{wform}
	W=fX + \frac{1}{2}\epsilon_X X^2.
\end{equation}

In this paper we consider metastable vacua and R-symmetry breaking in O'Raifeartaigh models from the perspective of the effective K\"ahler potential.  The ideas in this paper are not new.  Instead, our interest is to translate known ideas, often presented from the perspective of the superpotential, to the effective K\"ahler potential.  This will be done by expanding (\ref{kp1l}) in powers of $X$ and focusing on individual terms.  We will take this up in the next section.  In section \ref{formulas} we construct convenient formulas for computing each term in the expansion that do not require diagonalizing the mass matrix, which can be difficult if there are a large number of fields, and apply them to specific examples in section \ref{examples}.  In section \ref{sugra} we discuss motivation and applications for this work, in particular with supergravity.

Throughout we make no assumption of R-symmetry in the superpotential and allow for arbitrary R-charge assignments, so that both explicit and spontaneous R-symmetry breaking are included in our analysis.  In analyzing spontaneous symmetry  breaking the K\"ahler potential must be considered through order $X^6$.  We will write down general formulas, but when possible will give specific formulas through order $X^6$ in the K\"ahler potential.  We will also carry R-symmetry violating parameters, such as $\epsilon_X$ in (\ref{orsp2}), only through first order since they will always be assumed small.

%================================================================================================================

\section{The Effective K\"ahler Potential}
\label{ekp}

%------------------------------------------------------------------------------------------------------

\subsection{R-Symmetry Violating Terms}
\label{rsymterms}

The first question we might ask is which terms do we expect to show up in the effective K\"ahler potential?  If the superpotential is R-symmetric then the one loop correction (\ref{kp1l}) can only contain R-symmetric terms.  In general, upon expanding in $X$, it will contain all of them and be of the form
\begin{equation} \label{kp1lrsym}
	K_\tn{1-loop} = k_{0,0} + k_{2,1}|X|^2 + k_{4,2}|X|^4 + k_{6,3}|X|^6 + \cdots
\end{equation}
(the index notation will be explained in a moment).  If instead the superpotential is not R-symmetric, containing R-symmetry violating terms, so that we are using superpotential (\ref{orsp2}), then the one loop correction (\ref{kp1l}) will contain both R-symmetric and R-symmetry violating terms.  Upon expanding in $X$, the most general form is
\begin{equation} \label{kp1lgen}
	K_\tn{1-loop} = \sum_{n=0}\sum_{p=0}^{n} k_{n,p} X^p\overline{X}^{n-p}, \qquad \bar{k}_{n,p} = k_{n,n-p}.
\end{equation}  
The index notation is that $n$ is the total number of factors of $X$ and $\overline{X}$ and $p$ is the total number of factors of $X$ (so that $n-p$ is the total number of factors of $\overline{X}$).  A bar denotes conjugation and the second formula follows from the requirement that the K\"ahler potential be real.

Determining which terms show up in the loop correction, when R-symmetry violating terms are included, is not difficult and can be done with a simple analysis of R-symmetry.  We will explain with an example.  Consider the superpotential
\begin{equation}
	W = fX + m\phi_1 \phi_2 + \frac{1}{2}h X\phi_1^2 + \frac{1}{2}\epsilon \phi_2^2.
\end{equation}
The first three terms make up the original O'Raifeartaigh model \cite{O'R}, which is R-symmetric under the assignments $R(X)=2$, $R(\phi_1)=0$ and $R(\phi_2)=2$.  The final term explicitly breaks the R-symmetry.

However, if we assign the R-charge $R(\epsilon)=-2$ then the superpotential is R-symmetric.  This is a useful trick because if the superpotential is R-symmetric the one loop contribution to the K\"ahler potential must also be R-symmetric.  We still use the general K\"ahler potential (\ref{kp1lgen}), but now, using the R-charge assignment for $\epsilon$, we know exactly which terms will show up: The K\"ahler potential will include all R-symmetric terms that can be formed from $R(X)=2$ and $R(\epsilon)=-2$.  Such terms may be determined with the help of
\begin{equation}
	R(k_{n,p}) = 2(n-2p).
\end{equation}
The K\"ahler potential is of the form
\begin{equation} \label{k1lex}
	K_{\tn{1-loop}} \sim \left(\epsilon X + \tn{c.c}\right) + |X|^2 + \left(\epsilon |X|^2X + \tn{c.c.}\right) + |X|^4 + \left(\epsilon|X|^4 X + \tn{c.c.} \right) + |X|^6 + \cdots,
\end{equation}
where $\epsilon$ was kept only through first order, as mentioned in the introduction. (We did not write down the coefficients, which requires computing the K\"ahler potential exactly, and will be taken up in the next section.)

This trick not only determines which terms show up in the effective K\"ahler potential, but also which R-symmetry violating terms we might like to include in the superpotential when model building.  For example, in (\ref{k1lex}) we dropped the $\epsilon^2|X|^2 X^2$ term, but if we wanted an $|X|^2 X^2$ term we could obtain one by having an R-symmetry violating parameter whose R-charge is $-4$.  This would happen if we included, say, an $X\phi_2^2$ term in the superpotential.

%------------------------------------------------------------------------------------------------------

\subsection{The Inverse K\"ahler Metric}

Our primary interest in the effective K\"ahler potential is for constructing the scalar potential,
\begin{equation}
	V = K^{X\overline{X}}_{\tn{eff}} |W_X|^2,
\end{equation}
where $K^{X\overline X}_{\tn{eff}}$ is the inverse K\"ahler metric.  Our goal in this section is to find its general form.

From now on we will write the effective K\"ahler potential as $K$, without the subscript ``eff."  The K\"ahler metric is given by
\begin{equation} \label{kmet}
	K_{X\overline{X}} = \frac{\partial}{\partial X}\frac{\partial}{\partial \overline{X}}K = 1 + K^{\tn{1-loop}}_{X\overline X} = (1 + k_{2,1}) + \widetilde{K}^{\tn{1-loop}}_{X\overline X},
\end{equation}
where in the final form we separated out $k_{2,1}$ to define
\begin{equation}
	\widetilde{K}^{\tn{1-loop}}_{X\overline X} \equiv K^{\tn{1-loop}}_{X\overline X} - k_{2,1} = \sum_{n=3}\sum_{p=1}^{n-1} p(n-p)X^{p-1}\overline{X}^{n-p-1},
\end{equation}
with $K_\tn{1-loop}$ being given in (\ref{kp1lgen}).  The 1, on the right hand side of (\ref{kmet}), comes from the tree level contribution to the effective K\"ahler potential (\ref{effk}).  $k_{2,1}$ is loop suppressed and negligible compared to 1 and will be ignored.  This simplifies the inverse K\"ahler metric and was the reason for separating it out.

The inverse K\"ahler metric is given by
\begin{equation}
	K^{X\overline{X}} = \frac{1}{K_{X{\overline X}}} = 1-\widetilde{K}^{\tn{1-loop}}_{X\overline X} + \left(\widetilde{K}^{\tn{1-loop}}_{X\overline X}\right)^2 - \cdots.
\end{equation}
As mentioned in the introduction we will give explicit formulas through order $X^6$ in the effective K\"ahler potential.  Since the K\"ahler metric requires two derivatives this is equivalent to order $X^4$ in the K\"ahler metric and its inverse.  Also as mentioned in the introduction we will include R-symmetry violating terms only through first order.  The R-symmetric parameters are the $k_{n,n/2}$ (for even $n$) and the rest are R-symmetry violating.  The inverse K\"ahler metric to this order is
\begin{equation} \label{invkm}
\begin{aligned}
	K^{X\overline X} = 1 &- \left(2k_{3,2}X + \tn{c.c.}\right) \\
	&-4k_{4,2}\left|X\right|^2 - \left(3k_{4,3}X^2 + \tn{c.c.}\right) \\
	&-\left[4k_{5,4}X^3 + \left(6k_{5,3}-16k_{3,2}k_{4,2}\right)\left|X\right|^2X + \tn{c.c.}\right] \\
	&-\left(9k_{6,3}-16k_{4,2}^2\right)\left|X\right|^4 - \left[5k_{6,5}X^4 + \left(8k_{6,4}-24k_{4,2}k_{4,3}\right)\left|X\right|^2X^2 +\tn{c.c.}\right].
\end{aligned}
\end{equation}

%------------------------------------------------------------------------------------------------------

\subsection{The Scalar Potential and R-Symmetry Breaking}
\label{thespot}

The specific form of the scalar potential through order $X^4$ can be obtained using (\ref{invkm}), (\ref{spform}) and (\ref{wform}).  Including R-symmetry violating terms only through first order leads to
\begin{equation} \label{spot}
\begin{aligned}
	V = |f|^2 &+ \left[\left(\bar{f}\epsilon_X - 2|f|^2k_{3,2}\right)X + \tn{c.c.}\right] \\
	&-4|f|^2k_{4,2}|X|^2 - \left(3|f|^2k_{4,3}X^2 + \tn{c.c.}\right) \\
	&-\left[4|f|^2k_{5,4}X^3 + \left(6|f|^2k_{5,3} + 4\bar{f}\epsilon_X k_{4,2} - 16|f|^2k_{3,2}k_{4,2}\right)|X|^2X + \tn{c.c}\right] \\
	&+|f|^2\left(16k_{4,2}^2-9k_{6,3}\right)|X|^4 + |f|^2\left[\left(24k_{4,3}k_{4,2}-8k_{6,4}\right)|X|^2X^2 -5k_{6,5}X^4 +\tn{c.c.}\right].
\end{aligned}
\end{equation}
If $k_{4,2}<0$ there is a stable, R-symmetry breaking vacuum at
\begin{equation} \label{susyvac}
	X\approx \frac{2|f|^2\bar{k}_{3,2}-f\bar{\epsilon}_X}{4|f|^2|k_{4,2}|},
\end{equation}
assuming the R-symmetry violating terms are sufficiently small such that we can trust our expansion, with vacuum energy $V\approx |f|^2$.

If the superpotential is R-symmetric, then only the $k_{n,n/2}$ (for even $n$) are nonzero.  The scalar potential in this case is
\begin{equation} \label{rsymsp}
\begin{aligned}
	V &= |f|^2 -4|f|^2k_{4,2}|X|^2 + |f|^2\left(16k_{4,2}^2-9k_{6,3}\right)|X|^4 \\
	&\approx |f|^2 -4|f|^2k_{4,2}|X|^2 -9|f|^2k_{6,3}|X|^4,
\end{aligned}
\end{equation}
since $k_{4,2}^2$ is loop suppressed compared to $k_{6,3}$.  Spontaneous R-symmetry breaking will occur if
\begin{equation}\label{rsymsp2}
	k_{4,2} > 0 \quad \tn{and} \quad k_{6,3} < 0,
\end{equation}
with a stable, R-symmetry breaking vacuum at
\begin{equation}\label{rsymsp3}
	|X|^2=\frac{2k_{4,2}}{16k_{4,2}^2-9k_{6,3}} \approx \frac{2k_{4,2}}{9|k_{6,3}|},
\end{equation}
assuming $|X|$ is sufficiently small such that we can trust our expansion,
with vacuum energy $V\approx |f|^2$.
%=================================================================================================================

\section{Formulas for the Effective K\"ahler Potential}
\label{formulas}

In the previous section we wrote the expanded effective K\"ahler potential as
\begin{equation} \label{kp2lgen}
	K_\tn{1-loop} = \sum_{n=0}\sum_{p=0}^{n} k_{n,p} X^p\overline{X}^{n-p}, \qquad \bar{k}_{n,p} = k_{n,n-p}.
\end{equation}  
In this section, convenient formulas will be constructed for computing the $k_{n,p}$.  In doing this, we  make no assumption of R-symmetry and allow arbitrary R-charge assignments, so that both explicit and spontaneous R-symmetry breaking may be included.

%--------------------------------------------------------------------------------------------

\subsection{Expansion}

The one loop correction to the K\"ahler potential is given by \cite{effK, ISS1}
\begin{equation} \label{ekp2}
	K_{\tn{1-loop}} = -\frac{1}{32\pi^2}\tn{Tr}\left[\mathcal{M}^\dagger \mathcal{M}\ln\left(\frac{\mathcal{M}^\dagger \mathcal{M}}{\Lambda^2}\right)\right],
\end{equation}
where
\begin{equation}
	\mathcal{M}_{ij} = M_{ij} + XN_{ij}
\end{equation}
is the mass matrix.  To be definite, let $M_{ij}$ and $N_{ij}$ be $N_{\phi}\times N_{\phi}$ matrices so that the number of $\phi_i$ fields in the superpotential is $N_{\phi}$.  We would like to expand the above formula in powers of $X$ and $\overline{X}$, which is more easily accomplished from its integral representation \cite{ISS1}:
\begin{equation}\label{intexp}
	K_\tn{1-loop} = \frac{\Lambda}{32\pi^2}N_\phi - \frac{1}{16\pi^2}\tn{Tr}\int_0^\Lambda dv\, v^3 \left(v^2 + {\cal M}^\dagger{\cal M}\right)^{-1}.
\end{equation}
This formula is only valid in the limit $\Lambda \rightarrow \infty$, such that
\begin{equation}
	\frac{a}{b+\Lambda} \rightarrow 0, \qquad \ln\left(\frac{a^2}{b^2+\Lambda^2}\right) \rightarrow \ln\left(\frac{a^2}{\Lambda^2}\right), \qquad \tn{etc.,}
\end{equation}
where $a$ and $b$ are arbitrary, $\Lambda$-independent terms.

In the integral (\ref{intexp}) the term to be expanded is
\begin{equation}
	\left(v^2+{\cal M}^\dagger{\cal M}\right)^{-1} = \left[v^2 + M^\dagger M + |X|^2 N^\dagger N + XM^\dagger N + \overline{X}N^\dagger M\right]^{-1}.
\end{equation}
The expansion is defined by
\begin{equation} \label{expansion}
	\left[v^2 + M^\dagger M + |X|^2 N^\dagger N + XM^\dagger N + \overline{X}N^\dagger M\right]^{-1} = \sum_{n=0} \sum_{p=0}^{n} C_{n,p} X^q \overline{X}^{n-p},
\end{equation}
where the $C_{n,p}$ are matrices.  Multiplying both sides by the inverse of the left hand side and then reordering we arrive at
\begin{equation}
	1 = \sum_{n=0}\sum_{p=0}^n X^{p}\overline{X}^{n-p}\left[C_{n,p}\left(v^2 + M^\dagger M\right) + C_{n-2,p-1}N^\dagger N + C_{n-p,p-1}M^\dagger N + C_{n-1,p}N^\dagger M\right].
\end{equation}
Matching coefficients for like powers of $X$ allow us to solve for the $C_{n,p}$.

To write down the $C_{n,p}$ it is convenient to define
\begin{equation}\label{bdedef}
	b \equiv \left(v^2 + M^{\dagger}M\right)^{-1} N^\dagger N, \quad d \equiv \left(v^2 + M^{\dagger}M\right)^{-1} M^{\dagger}N, \quad e \equiv \left(v^2 + M^{\dagger}M\right)^{-1} N^{\dagger}M.
\end{equation}
We will also make use of the symbol Com, which stands for combination, and gives the sum of all possible orderings of distinct elements.  For example,
\begin{equation}
	\tn{Com}(d^2e^2) = ddee + dede+deed + edde+eded+eedd.
\end{equation}
With these definitions, the $C_{n,p}$ are
\begin{equation}
\begin{aligned}
	C_{n,p} &= (-1)^n \tn{Com}\left[ d^p e^{n-p} - d^{p-1}be^{n-p-1} +d^{p-2}b^2e^{n-p-2} - \cdots +(-1)^{n-p}d^{2p-n}\right]\left(v^2 + M^\dagger M\right)^{-1} \\
	&=(-1)^n\sum_{\ell=0}^{n-p} (-1)^\ell \tn{Com}\left(d^{p-\ell}b^\ell e^{m-p-\ell}\right)\left(v^2 + M^\dagger M\right)^{-1},
\end{aligned}
\end{equation}
which is valid for $p\geq n/2$ with the other values given by $C_{n,p} = C^{\dagger}_{n,n-p}$

The $C_{n,p}$ are to be placed in the integral (\ref{intexp}) through their definition (\ref{expansion}).  As we are after the $k_{n,p}$ in (\ref{kp2lgen}), we find, for our final result,
\begin{equation} \label{knpgen}
	k_{n,p} = \delta_{n,0} \frac{\Lambda}{32\pi^2}N_\phi - \frac{(-1)^n}{16\pi^2}\tn{Tr}\int_0^\Lambda dv \, v^3 \left[\sum_{\ell=0}^{n-p} (-1)^\ell \tn{Com}\left(d^{p-\ell}b^\ell e^{m-p-\ell}\right)\right]\left(v^2 + M^\dagger M\right)^{-1},
\end{equation}
which is valid for $p\geq n/2$ with the other values given by $	k_{n,p} = \bar{k}_{n,n-p}$.

%---------------------------------------------------------------------------------------------------------------------

\subsection{Formulas}

The combinations in (\ref{knpgen}) are cumbersome.  Fortunately they can be removed.  Each of the $k_{n,p}$ can be written as a full derivative.  An integration by parts can then be used, simplifying the formulas and removing the combinations.  As mentioned in the introduction we will give specific formulas through order $X^6$, so for $n\leq 6$:
{\allowdisplaybreaks
\begin{subequations}\label{ks}
\begin{align}
	k_{0,0} &= \frac{\Lambda}{32\pi^2}N_\phi - \frac{1}{16\pi^2}\tn{Tr}\int_0^\Lambda dv \, v^3\left(v^2 + M^\dagger M\right)^{-1} \\
	k_{1,1} &=-\frac{1}{32\pi^2}\tn{Tr}\left(N^\dagger M\right) + \frac{1}{16\pi^2} \tn{Tr}\int_0^\Lambda dv \, v (d) \\
	k_{2,2} &=-\frac{1}{16\pi^2}\tn{Tr} \int_0^\Lambda dv \, v \left(\frac{1}{2}d^2\right) \\
	k_{2,1} &=-\frac{1}{32\pi^2}\tn{Tr}\left(N^\dagger N\right) - \frac{1}{16\pi^2}\tn{Tr}\int_0^\Lambda dv \, v(de-b)\\
	k_{3,3} &= \frac{1}{16\pi^2}\tn{Tr} \int_0^\Lambda dv \, v \left(\frac{1}{3}d^3\right) \\
	k_{3,2} &= \frac{1}{16\pi^2}\tn{Tr} \int_0^\Lambda dv \, v (d^2e - db) \\
	k_{4,4} &= -\frac{1}{16\pi^2}\tn{Tr} \int_0^\Lambda dv \, v \left(\frac{1}{4}d^4\right) \\
	k_{4,3} &= -\frac{1}{16\pi^2}\tn{Tr} \int_0^\Lambda dv \, v (d^3 e - d^2 b) \\
	k_{4,2} &= -\frac{1}{16\pi^2}\tn{Tr} \int_0^\Lambda dv \, v \left[d^2 e^2+\frac{1}{2}(de)^2 -deb-dbe+\frac{1}{2}b^2\right] \\
	k_{5,5} &= \frac{1}{16\pi^2}\tn{Tr} \int_0^\Lambda dv \, v \left(\frac{1}{5}d^5\right) \\
	k_{5,4} &= \frac{1}{16\pi^2}\tn{Tr} \int_0^\Lambda dv \, v (d^4 e - d^3 b) \\
	k_{5,3} &= \frac{1}{16\pi^2}\tn{Tr} \int_0^\Lambda dv \, v \left[d^3e^2+d(de)^2-d^2 be -d^2eb-dbde+db^2\right] \\
	k_{6,6} &= -\frac{1}{16\pi^2}\tn{Tr} \int_0^\Lambda dv \, v \left(\frac{1}{6}d^6\right) \\
	k_{6,5} &= -\frac{1}{16\pi^2}\tn{Tr} \int_0^\Lambda dv \, v \left(d^5e - d^4b\right) \\
	k_{6,4} &= -\frac{1}{16\pi^2}\tn{Tr} \int_0^\Lambda dv \, v \biggl[d^4e^2+d^3ede+\frac{1}{2}(d^2e)^2-d^3be-d^3eb-d^2bde-d^2edb \notag\\
	&\qquad\qquad\qquad\qquad+d^2b^2+\frac{1}{2}(db)^2\biggr] \\
	k_{6,3} &= -\frac{1}{16\pi^2}\tn{Tr} \int_0^\Lambda dv \, v \biggl[d^3e^3 +ed(de)^2 + de(ed)^2 + \frac{1}{3}(de)^3 - d^2be^2-d^2ebe-b(ed)^2 \notag \\
	&\qquad\qquad\qquad\qquad-b(de)^2-de^2db-d^2e^2b+deb^2+db^2e+dbeb-\frac{1}{3}b^3\biggr].
\end{align}
\end{subequations}
}The first terms in $k_{1,1}$ and $k_{2,1}$ are nonvanishing boundary terms from the integration by parts.  The remaining $k_{n,p}$ are obtained from $k_{n,p} = \bar{k}_{n,n-p}$.

%---------------------------------------------------------------------------------------------------------------------

\subsection{R-Symmetric Terms}

The R-symmetric terms ($k_{2,1}$, $k_{4,2}$ and $k_{6,3}$) in (\ref{ks}) appear comparatively more complicated.  There exists another formulation that leads to simpler formulas for the R-symmetric terms which we develop here.  

\subsubsection{Setup}

As in the previous section, the one loop correction to the K\"ahler potential is given by \cite{effK, ISS1}
\begin{equation} \label{kp1l2}
	K_{\tn{1-loop}} = -\frac{1}{32\pi^2}\tn{Tr}\left[\mathcal{M}^\dagger \mathcal{M}\ln\left(\frac{\mathcal{M}^\dagger \mathcal{M}}{\Lambda^2}\right)\right],
\end{equation}
where
\begin{equation}
	\mathcal{M}_{ij} = M_{ij} + XN_{ij}
\end{equation}
is the mass matrix.  We again take $M_{ij}$ and $N_{ij}$ to be $N_{\phi}\times N_{\phi}$ matrices.  Our departure from the previous subsection begins by removing the Hermitian conjugate.  This can be done by defining the $2N_{\phi} \times 2N_{\phi}$ matrices
\begin{equation} \label{matrixdef}
	\widehat{M} \equiv \left[ {\begin{array}{*{20}c}
  	0 & {M^\dagger }  \\
  	M & 0  \\
	\end{array} } \right], \quad
	\widehat{N} \equiv \left[ {\begin{array}{*{20}c}
  	0 & {N^\dagger }  \\
  	N & 0  \\
	\end{array} } \right], \quad
	\widehat{X} \equiv \left[ {\begin{array}{*{20}c}
  	{\overline{X} I} & 0  \\
  	0 & {XI}  \\
	\end{array} } \right]
\end{equation} 
($I$ is the $N_{\phi}\times N_{\phi}$ identity matrix) and
\begin{equation}
	\widehat{\mathcal{M}} \equiv \widehat{M}+\widehat{X}\widehat{N}.
\end{equation}
Now,
\begin{equation}
	\widehat{\mathcal{M}}^2=\left[ {\begin{array}{*{20}c}
  	\mathcal{M^\dagger M} & 0  \\
  	0 & \mathcal{MM^\dagger}  \\
	\end{array} } \right]
	=\left[ {\begin{array}{*{20}c}
  	\left(\mathcal{MM^\dagger}\right)^* & 0  \\
  	0 & \mathcal{MM^\dagger}  \\
	\end{array} } \right],
\end{equation}
the second matrix following from $\cal{M}$ being symmetric.  Since $\mathcal{MM^\dagger}$ has real eigenvalues (since it is Hermitian), $\widehat{\mathcal{M}}^2$ has two copies of each of the eigenvalues of $\mathcal{M}^\dagger \mathcal{M}$.  Thus
\begin{equation}\label{kp1l3}
	K_{\tn{1-loop}} = -\frac{1}{64\pi^2}\tn{Tr}\left[\widehat{\mathcal{M}}^2\ln\left(\frac{\widehat{\mathcal{M}}^2}{\Lambda^2}\right)\right].
\end{equation}
The extra factor of 1/2 is to correct for doubling the eigenvalues.

The expansion will be performed using the integral representation of (\ref{kp1l3}) \cite{ISS1}:
\begin{equation} \label{kpint}
	K_{\tn{1-loop}} = \frac{\Lambda^2}{32\pi^2}N_\phi-\frac{1}{32\pi^2} \tn{Tr} \int_0^\Lambda dv \, v^3 \left(v^2+\widehat{\mathcal{M}}^2\right)^{-1}.
\end{equation}
This formula is only valid in the limit $\Lambda\rightarrow\infty$, just as was (\ref{intexp}).

%---------------------------------------------------------------------------------------------------------------------

\subsubsection{Expansion}

Our interest here is only with the R-symmetric terms (i.e. the $k_{n,n/2}$ for even $n$).  These terms enter the one loop correction to the K\"ahler potential as
\begin{equation} \label{k1lrsym}
	K_\tn{1-loop} = k_{0,0} + k_{2,1}|X|^2 + k_{4,2}|X|^4 + k_{6,3}|X|^6+\cdots.
\end{equation}
This equation tells us that we may make the simplifying assumption that $X$ is real \cite{Shih} and hence, from (\ref{matrixdef}), that $\widehat{X}=X$ is proportional to the identity.  The $\widehat{X}$'s locked up inside the integral (\ref{kpint}), now proportional to the identity, may be pulled out, simplifying expressions.

In (\ref{kpint}) the term to be expanded is
\begin{equation}
	\left(v^2+\widehat{\mathcal{M}}^2\right)^{-1} = \left[\left(v^2 + \widehat{M}^2\right) + X\left\{\widehat{M},\widehat{N}\right\} + X^2 \widehat{N}^2\right]^{-1},
\end{equation}
where we have assumed $X$ is real.  The expansion is defined by
\begin{equation} \label{cndef}
	\left[\left(v^2 + \widehat{M}^2\right) + X\left\{\widehat{M},\widehat{N}\right\} + X^2 \widehat{N}^2\right]^{-1} = \sum\nolimits_n X^n C_n,
\end{equation}
where the $C_n$ are matrices.  Multiplying both sides by the inverse of the left hand side and then matching coefficients of like powers of $X$ will allow us to determine the $C_n$.  

To write down the $C_n$ it is convenient to define
\begin{equation} \label{ABdef}
	A \equiv \left(v^2 + \widehat{M}^2\right)^{-1}\left\{\widehat{M},\widehat{N}\right\}, \qquad B \equiv \left(v^2 + \widehat{M}^2\right)^{-1}\widehat{N}^2,
\end{equation}
and, as before, make use of Com, which gives the sum of all possible orderings of distinct elements.  For example,
\begin{equation}
	\tn{Com}\left(AABB\right)=AABB+ABAB+BAAB+ABBA+BABA+BBAA.
\end{equation}
With these definitions, the $C_n$ in the expansion (\ref{cndef}) are
\begin{equation}
\begin{aligned}
	C_n &= (-1)^n \tn{Com}\left[ A^n - A^{n-2}B + A^{n-4}B^2 +\cdots + 
	\left\{ {\begin{array}{*{20}l}
   {A^{n/2} } & \tn{for even $n$}  \\
   {AB^{(n - 1)/2} } & \tn{for odd $n$}  \\
	\end{array} } \right\}
	\right]\left(v^2 + \widehat{M}^2\right)^{-1} \\ 
	& = (-1)^n \left[\sum\nolimits_\ell (-1)^\ell \tn{Com} \left(A^{n-2\ell}B^\ell\right)\right]\left(v^2 + \widehat{M}^2\right)^{-1},
\end{aligned}
\end{equation}
where $\ell$ goes from zero to $n/2$ for even $n$ and from zero to $(n-1)/2$ for odd $n$.  

The $C_n$ are to be placed in the integral (\ref{kpint}) though their definition (\ref{cndef}).  From (\ref{k1lrsym}) we can then write down a general expression for the $k_{n,n/2}$:
\begin{equation} \label{kpint3}
	k_{n,n/2} = \delta_{n,0}\frac{\Lambda^2}{32\pi^2}N_{\phi} -\frac{1}{32\pi^2} \tn{Tr} \int_0^\Lambda dv \, v^3 \left[\sum_{\ell=0}^{n/2} (-1)^\ell \tn{Com} \left(A^{2n-2\ell}B^\ell\right)\right]\left(v^2 + \widehat{M}^2\right)^{-1}.
\end{equation}
While this expression was derived with an R-symmetric superpotential in mind, it works equally well when the superpotential contains R-symmetry violating terms.  The reason is that we are limiting our equations to first order in R-symmetry violating terms and the earliest they enter the $k_{n,n/2}$ is at second order.

%-----------------------------------------------------------------------------------------------------------------------------

\subsubsection{Formulas}

As in the general case above, the combinations in (\ref{kpint3}) can be removed by rewriting them as full derivatives and using an integration by parts.  Specific formulas for $n\leq 6$ are
\begin{subequations}\label{rsymform}
\begin{align}
	k_{2,1} &= \frac{1}{2} \tn{Tr} \left(\widehat{N}^2\right) - \frac{1}{32\pi^2} \tn{Tr} \int_0^\Lambda dv \, v \left(\frac{1}{2}A^2-B\right) \\
	k_{4,2} &= -\frac{1}{32\pi^2} \tn{Tr} \int_0^\Lambda dv \, v \left(\frac{1}{4}A^4-A^2B + \frac{1}{2}B^2\right) \label{k42int}\\
	k_{6,3} &= -\frac{1}{32\pi^2} \tn{Tr} \int_0^\Lambda dv \, v \left(\frac{1}{6}A^6-A^4B + A^2B^2 +\frac{1}{2}(AB)^2 - \frac{1}{3}B^3\right),
\end{align}
\end{subequations}
where the first term in $k_{2,1}$ is a nonvanishing boundary term from the integration by parts.  These formulas are simpler than their counterparts in (\ref{ks}).

%==================================================================================================

\section{Examples}
\label{examples}

The formulas developed in section \ref{formulas} will now be applied to two examples.  The first is the original O'Raifeartaigh model \cite{O'R} with explicit R-symmetry breaking terms added.  The second is Shih's model \cite{Shih} which exhibits spontaneous R-symmetry breaking.  While in both cases the effective K\"ahler potential may be computed directly from (\ref{kp1l}), these examples offer a straightforward illustration of the use of our formulas.

%--------------------------------------------------------------------------------------------------

\subsection{Explicit R-Symmetry Breaking: The O'Raifeartaigh Model}

The original O'Raifeartaigh Model has superpotential \cite{O'R}
\begin{equation}
	W =  fX + m\phi_1\phi_2 + \frac{1}{2} h X\phi_1^2,
\end{equation}
which is R-symmetric under R-charge assignments
\begin{equation}
	R(X)=2, \quad R(\phi_1)=0, \quad R(\phi_2)=2.
\end{equation}
We can add explicit R-symmetry violating terms:
\begin{equation} \label{orbr}
	W = fX +\frac{1}{2}\epsilon_X X^2 + \frac{1}{2}\left[ \left( 2m\phi_1\phi_2 + \epsilon_1 \phi_1^2 + \epsilon_2 \phi_2^2 \right) + X\left( h\phi_1^2 + 2\delta_{12} \phi_1\phi_2 + \delta_{22}\phi_2^2\right)\right].
\end{equation}
R-symmetry may be restored in (\ref{orbr}) with R-charge assignments
\begin{equation} \label{orrca}
	R(\epsilon_X)=R(\epsilon_2)=R(\delta_{12})=-2, \quad R(\epsilon_1)=2, \quad R(\delta_{22})=-4.
\end{equation}
For simplicity, we assume all parameters are real.

To compute the coefficients in the expanded effective K\"ahler potential (the $k_{n,p}$ in (\ref{kp1lgen})), we first construct, from (\ref{orsp2}) and (\ref{orbr}), the matrices
\begin{equation}
M = \left[ {\begin{array}{*{20}c}
   {\epsilon _1 } & m  \\
   m & {\epsilon _2 }  \\
 \end{array} } \right], \qquad
N = \left[ {\begin{array}{*{20}c}
   h  & {\delta _{12} }  \\
   {\delta _{12} } & \delta_{22}  \\
 \end{array} } \right].
\end{equation}
With these we construct the matrices (\ref{bdedef}):
{\allowdisplaybreaks
\begin{subequations}
\begin{align}
b &= \frac{h}{v^2 + m^2}\left[ {\begin{array}{*{20}c}
   h & \delta _{12} \\
 \frac{\left(m^2+v^2\right) \delta _{12}-h m \left(\epsilon _1+\epsilon _2\right)}{m^2+v^2} & 0
 \end{array} } \right] \\
d &=\frac{1}{v^2 + m^2}\left[ {\begin{array}{*{20}c}
   \frac{m \left(m^2+v^2\right) \delta _{12}+h \left(v^2 \epsilon _1-m^2 \epsilon _2\right)}{m^2+v^2} & m \delta _{22} \\
  mh & m \delta _{12}
 \end{array} } \right] \\
e &=\frac{1}{v^2 + m^2}\left[ {\begin{array}{*{20}c}
  m \delta _{12}+h \epsilon _1 &  m h \\
 m \delta _{22} & \frac{m \left[\left(m^2+v^2\right) \delta _{12}-h m \left(\epsilon _1+\epsilon _2\right)\right]}{m^2+v^2}
 \end{array} } \right],
\end{align}
\end{subequations}
}which make up our formulas (\ref{ks}) for the $k_{n,p}$.  We find
{\allowdisplaybreaks
\begin{subequations}\label{orks}
\begin{align}
	k_{3,2} &= \frac{1}{32\pi^2}\left(\frac{h^3}{6m^2}\epsilon_2 - \frac{h^3}{3m^2}\epsilon_1 - \frac{h^2}{m}\delta_{12}\right) \\
	k_{4,3} &=\frac{1}{32\pi^2}\frac{h^3}{6m^2}\delta_{22}\\
	k_{4,2} &= -\frac{1}{32\pi^2} \frac{h^4}{6m^2} \\
	k_{5,3} &= -\frac{1}{32\pi^2}\left(\frac{h^5}{30m^4}\epsilon_2 - \frac{h^5}{20m^4}\epsilon_1 - \frac{h^4}{6m^3}\delta_{12}\right) \\
	k_{6,4} &= -\frac{1}{32\pi^2}\frac{h^5}{30m^4}\delta_{22} \\
	k_{6,3} &= \frac{1}{32\pi^2} \frac{h^6}{60m^4}
\end{align}
\end{subequations}
}for the nonzero $k_{n,p}$ that show up in the scalar potential (the remaining terms are obtained from $k_{n,p} = \bar{k}_{n,n-p}$).  These are precisely the terms we expected to find based on the discussion in section \ref{rsymterms} and the R-charge assignments (\ref{orrca}).

The scalar potential is given by
\begin{equation}
	V = K^{X\overline{X}} \left| W_X \right|,
\end{equation}
where now
\begin{equation}
	W = fX + \frac{1}{2}\epsilon_X X^2.
\end{equation}
The general form was written down in (\ref{spot}).  Plugging the $k_{n,p}$ in (\ref{orks}) into (\ref{spot}) would give us the looped corrected scalar potential.

As discussed in section \ref{thespot}, since $k_{4,2}<0$ a stable, R-symmetry breaking vacuum, given by (\ref{susyvac}), exists at
\begin{equation}
	X \approx 3\frac{m^2}{h^4}\left( \frac{h^3}{6m^2}\epsilon_2 - \frac{h^3}{3m^2}\epsilon_1 - \frac{h^2}{m}\delta_{12} - 16\pi^2\frac{\epsilon_X}{f}\right),
\end{equation}
assuming the R-symmetry violating terms are sufficiently small such that we can trust our expansion, with $V\approx f^2$.

%--------------------------------------------------------------------------------------------------

\subsection{Spontaneous R-Symmetry Breaking: Shih's Model}

Shih proved that in the R-symmetric superpotential 
\begin{equation} \label{orsp3}
	W=fX + \frac{1}{2}\left(M_{ij}+XN_{ij}\right)\phi_i\phi_j,
\end{equation}
spontaneous R-symmetry breaking can only occur if there are fields with R-charge assignments different from zero and two \cite{Shih}.  The simplest possible superpotential exhibiting spontaneous R-symmetry breaking is \cite{Shih}
\begin{equation}\label{shihsp}
	W=fX + \lambda X \phi_1 \phi_2 + m_1 \phi_1\phi_3 + \frac{1}{2}m_2\phi_2^2,
\end{equation}
with R-charge assignments
\begin{equation}
	R(X) = 2, \quad R(\phi_1)=-1, \quad R(\phi_2)=1, \quad R(\phi_3)=3.
\end{equation}
As in the previous example, for simplicity, we take all parameters to be real.

Since the superpotential is R-symmetric, the effective K\"ahler potential will be R-symmetric as well and be of the form
\begin{equation}
	K_\tn{1-loop} = k_{0,0} + k_{2,1}|X|^2 + k_{4,2}|X|^4 + k_{6,3}|X|^6+\cdots.
\end{equation}
In other words, it will only contain the $k_{n,n/2}$ (for even $n$).  To compute these coefficients we start by comparing (\ref{shihsp}) with (\ref{orsp3}) to construct the matrices
\begin{equation}
M = \left[ {\begin{array}{*{20}c}
   0 & 0 & {m_1 }  \\
   0 & {m_2 } & 0  \\
   {m_1 } & 0 & 0  \\ \end{array} } \right], \qquad
N = \left[ {\begin{array}{*{20}c}
   0 & \lambda  & 0  \\
   \lambda  & 0 & 0  \\
   0 & 0 & 0  \\ \end{array} } \right].
\end{equation}
In this example we'll use the R-symmetric formulas (\ref{rsymform}).  For convenience we define
\begin{equation}
	z_1 \equiv \frac{1}{v^2 + m_1^2}, \qquad z_2 \equiv \frac{1}{v^2+m_2^2},
\end{equation}
so that we can more compactly write the matrices in (\ref{ABdef}) (themselves coming from the matrices in (\ref{matrixdef})):
\begin{equation}
	A = \lambda\left[
\begin{array}{cccccc}
 0 &    z_1m_2 & 0 & 0 & 0 & 0 \\
    z_2m_2 & 0 &    z_2 m_1& 0 & 0 & 0 \\
 0 &    z_1m_1 & 0 & 0 & 0 & 0 \\
 0 & 0 & 0 & 0 &    z_1 m_2& 0 \\
 0 & 0 & 0 &   z_2 m_2  & 0 &    z_2m_1 \\
 0 & 0 & 0 & 0 &    z_1m_1 & 0
\end{array}
\right], \quad
B= \lambda^2 \tn{diag}\left(z_1,z_2,0,z_1,z_2,0\right).
\end{equation}
Further defining
\begin{equation}
	r\equiv \frac{m_2}{m_1},
\end{equation}
then from our formulas (\ref{rsymform}) we find \cite{lpt}
\begin{subequations}
\begin{align}
k_{4,2} &= \frac{\lambda^4}{32\pi^2 m_1^2} \frac{1+2r^2-3r^4+r^2(3+r^2)\ln (r^2)}{(r^2-1)^3} \\
k_{6,3} &= -\frac{\lambda^6}{32\pi^2m_1^4} \frac{1+27r^2-9r^4-19r^6+6r^2(2+5r^2+r^4)\ln(r^2)}{3(r^2-1)^5}.
\end{align}
\end{subequations}
By plugging these equations into (\ref{rsymsp}) we arrive at the loop corrected scalar potential.

Now,
\begin{equation}
	k_{4,2} > 0 \quad \tn{for} \quad r>r_* \equiv 2.12,
\end{equation}
and since $k_{6,3}<0$ for $r>r_*$, then from (\ref{rsymsp2}) spontaneous R-symmetry breaking occurs for $r>r_*$ \cite{Shih,lpt} and the vacuum, from (\ref{rsymsp3}), is located at
\begin{equation}
	|X|^2\approx \frac{2k_{4,2}}{9|k_{6,3}|},
\end{equation}
assuming $|X|$ is sufficiently small such that we can trust our expansion,
with $V\approx f^2$.

%==================================================================================================

\section{Applications: Supergravity}
\label{sugra}

An important application for our formulas is supergravity, which modifies the scalar potential with Planck mass suppressed corrections.  It is interesting to consider the affect these corrections have on metastable vacua and R-symmetry breaking \cite{Kitano, ako, haba, lpt}.  The supergravity scalar potential is given by
\begin{equation}
	V=e^K \left(K^{X\overline{X}} \left|W_X + K_X W\right|^2 - 3\left|W\right|^2\right).
\end{equation}
An immediate consequence of including supergravity corrections, since the superpotential enters without derivatives, is a constant term in the superpotential surviving in the scalar potential.  Such a term is usually included for tuning the vacuum energy to zero, which sets its value.  However, a constant term explicitly breaks R-symmetry and may have important consequences for metastable vacua \cite{Kitano}.

Loop corrections in supergravity are complicated, but are important when analyzing the vacuum structure.  A simple way to include them is to use the effective K\"ahler potential \cite{Kitano, lpt}.  This was the original motivation for this paper and is currently under study.

%===ACKNOWLEDGEMENTS=============================================================================

\section*{Acknowledgments}

BK would to thank Linda M. Carpenter and Flip Tanedo for helpful discussions.  This work was supported in part by a New Jersey Space Grant Consortium fellowship.

%===REFERENCES============================================================================

\end{document}